\documentclass{jkas}

\def\year{2014} % publication year
\def\month{December} % publication month
\def\volume{00} % journal volume
\def\issue{0} % journal issue
\def\beginpage{1} % first page of article
\def\endpage{7} % last page of article
\def\received{October 22, 2014} % date paper was received by JKAS
\def\accepted{December 5, 2014} % date of acceptance
\date{Received \received; accepted \accepted}

\input colordvi

\usepackage{flushend} %% balance columns on last page
\title{
Intrinsic Brightness Temperatures of Compact Radio Jets\\ as a Function of Frequency
}

\author[1,2]{Sang-Sung Lee}
\affil[1]{Korea Astronomy and Space Science Institute, 776 Daedeokdae-ro,
Yuseong-gu, Daejeon 305-348, Korea; \email{sslee@kasi.re.kr}}
\affil[2]{Korea University of Science and Technology, 176 Gajeong-dong, Yuseong-gu, Daejeon 305-350, Korea}
\def\corrauthor{
S.-S. Lee
}

\def\runningauthor{
S.-S. Lee
}

\def\runningtitle{
Intrinsic Brightness Temperatures of Radio Jets as Function of Frequency
}

\def\keywords{
Galaxies: nuclei --- quasars: relativistic jets --- radio: galaxies
}

\def\abstracttext{
We present results of our investigation
of the radio intrinsic brightness temperatures of compact radio jets.
The intrinsic brightness temperatures of about 100 compact radio jets
at 2, 5, 8, 15, and 86~GHz
are estimated based on large VLBI surveys conducted in 2001-2003
(or in 1996 for the 5~GHz sample). 
The multi-frequency intrinsic brightness temperatures
of the sample of jets  
are determined by a statistical method relating
the observed brightness temperatures with the maximal apparent jet speeds,
assuming one representative intrinsic brightness temperature for a sample
of jets at each observing frequency.
By investigating the observed brightness temperatures at 15~GHz 
in multiple epochs, we found that the determination of
the intrinsic brightness temperature for our sample is affected by 
the flux density variability of individual jets at 
time scales of a few years.
This implies that it is important to use contemporaneous
VLBI observations for the multi-frequency analysis
of intrinsic brightness temperatures.
Since our analysis is based on the VLBI observations conducted in 2001-2003,
the results are not strongly affected by the flux density variability.
We found that the intrinsic brightness temperature $T_{\rm 0}$ increases
as $T_{\rm 0}\propto\nu_{\rm obs}^{\xi}$ with $\xi=0.7$ below
a critical frequency $\nu_{\rm c}\approx9~{\rm GHz}$
where energy losses begin to dominate the emission.
Above $\nu_{\rm c}$, $T_{\rm 0}$ decreases
with $\xi=-1.2$, supporting for
the decelerating jet model or particle cascade model.  
We also found that the peak value of $T_{\rm 0}\approx3.4\times10^{10}$~K
is close to the equipartition temperature, implying that
the VLBI cores observable at 2-86~GHz may be representing
jet regions where the magnetic field energy dominates
the total energy in jets.
}

\begin{document}
\jkashead %% set title, authors, abstract, etc.

%%%%%%%%%%%%%%%%%%%%%%%%%%%%%%%%%%%%%%%%%%%%%%%%%%%%%%%%%%%%%%%%%%%%%
%%% BEGIN MAIN TEXT HERE %%%%%%%%%%%%%%%%%%%%%%%%%%%%%%%%%%%%%%%%%%%%
%%%%%%%%%%%%%%%%%%%%%%%%%%%%%%%%%%%%%%%%%%%%%%%%%%%%%%%%%%%%%%%%%%%%%

\section{Introduction\label{sec:intro}}

Relativistic outflows (or jets) are ubiquitous phenomena associated with 
the most energetic and compact astronomical objects such as
GRB (Gamma-Ray Burst), AGN (Active Galactic Nucleus),
and XRB (X-Ray Binary or ``microquasar'') sources. 
We can gain deep insights into the processes
of the creation, the physics, and the behavior of relativistic jets
by studying the different or unified properties of
these objects. 
For this purpose, the relativistic jets in AGN
have been thoroughly studied
by many astronomers and astrophysicists  
both theoretically and observationally~\citep[e.g.,][]{BK79,mar95,lob98}.

Although a relativistic jet is quite a complicated phenomenon
and the underlying mechanisms of the formation,
collimation, acceleration, etc., of jets are far from being known,
we rely on knowledge about the relativistic jets obtained from
the theoretical and observational studies as follows.
Relativistic jets in AGN are formed in the immediate vicinity of 
the central black hole, 
and they interact with every major constituent of the AGN~\citep[see][]{LZ06}.
Relativistic jets are currently known to be driven from 
either the inner accretion disk~\citep{BP82} or the ergosphere of a rotating
black hole~\citep{BZ77}. 
The most upstream emission region
in the compact radio jet (the core) in Very Long Baseline Interferometry (VLBI)
observations does not represent
the origin (or launching site) of the jet.
The core is generally believed to represent 
a region of the jet where the optical depth at the observing frequency
is unity~\citep{kon81,lob98b}. 
In the region between the origin and the VLBI core of the jet,
the energy from the origin
is transferred to the core by a disturbance passing through it~\citep{MG85}. 
Jets are believed to be
collimated and accelerated by a twisted, most likely poloidal, magnetic
field~\citep[e.g.,][]{mei+01,jor+07}, which could be perturbed and generate 
the disturbance. For the relativistic jets of radio-loud galaxies,
magnetic fields play dynamically important roles in the collimation
and acceleration of the jets~\citep{zam+14}.

High-resolution VLBI observation
is one of the observational methods of studying the relativistic jets. 
The main observables of the relativistic jets with the VLBI observations
at radio wavelengths are the flux density and size of compact emission
regions, and hence the brightness temperatures.
Since, however, the emission from the relativistic jets are highly
Doppler-boosted, the observed brightness temperatures of the compact emission
regions do not represent the intrinsic properties of the relativistic jets.
In order to investigate the intrinsic properties, \cite{hom+06} applied 
a statistical method to estimate an intrinsic brightness temperature
for a sample of compact radio jets observed at 15~GHz.
They found that the derived intrinsic brightness temperature
for their sample is close to the equipartition temperature,
implying that the energies of particles and magnetic fields
in the emission regions are balanced.
However, \cite{lee13} applied the same method to the 86~GHz VLBI survey
observation data and found that the intrinsic brightness temperature
estimated for about 100 compact radio jets observed at 86~GHz
is lower than the 15~GHz intrinsic brightness temperature for the same sample
and, hence, the equipartition temperature.
This result indicates that the energy of the emission region observed
at 86~GHz is dominated by the magnetic field. 

In this paper, we investigate further
the intrinsic brightness temperatures of compact radio sources,
derived from multi-frequency large VLBI surveys.
In addition to the VLBI surveys at 15 and 86~GHz~\citep[see][and references therein]{lee13}, we also used
the VLBI surveys at 2, 5, and 8~GHz for this study. 
In Section 2, we describe the VLBI surveys at different frequencies,
show the effect of flux variability on the analysis,
and estimate the multi-frequency intrinsic brightness temperatures.  
In Section 3, we summarize what we found from the analysis.
In Section 4, we discuss our findings, and 
we present our conclusions in Section 5.

\section{Data and Analysis\label{data_and_analysis}}

\subsection{Data\label{data}}

In addition to the observed brightness temperatures at 15 and
86~GHz~\citep{lee+08,lis+09,lis+13},
we compiled the observed brightness temperatures
at lower frequencies of 2, 5, and 8~GHz from the large VLBI
surveys~\citep{sco+04,dod+08,PK12}.
\cite{PK12} used 19 global VLBI observing sessions conducted simultaneously
at 2.3~GHz and 8.6~GHz with up to 24 radio telescopes consisting of
10 Very Long Baseline Array (VLBA) stations and up to 14 additional
geodetic telescopes. The observations were carried out
in the period between 1998 October to 2003 September.
They observed a sample of 370 compact radio sources, including
251 quasars, 46 BL Lacertae objects,
31 radio galaxies, and 42 optically unidentified sources.
The naturally weighted images were fitted with circular Gaussian models
in order to derive physical properties of jet components.
Due to the high compactness of the VLBI core components,
they considered the core to be unresolved if its size, resulting from a fit with
a circular Gaussian model, was smaller than the minimum resolvable
size calculated following the same criteria as that described in \cite{lee13}.
The median values of the angular and projected linear sizes
of the VLBI core components are 1.04~mas (6.75~pc) and 0.28~mas (1.90~pc)
at 2.3~GHz and 8.6~GHz, respectively.
The dynamic range of the images is 106-4789 with a median of about 1000
at 2.3~GHz and 66-7042 with a median of about 1200 at 8.3~GHz.
The typical rms noise level is  about 0.5~mJy~beam$^{-1}$ at 2.3~GHz
and about 0.4~mJy~beam$^{-1}$ at 8.6~GHz. 
The observed brightness temperatures of the VLBI core components
are estimated based on the results of the imaging and Gaussian model-fitting.
The median values of the observed brightness temperatures for
the core components are $2.5\times10^{11}$~K
at both frequencies.

\cite{sco+04} and \cite{dod+08} detected and imaged 242 compact radio sources
with the VLBI Space Observatory Programme (VSOP) mission which
was a Japanese-led project using an orbiting 8~m telescope, HALCA,
along with global arrays of ground-based telescopes. 
The observations began in 1997 August and lasted until 2003 October.
The VSOP survey observations were done at 5~GHz and
the typical rms detection sensitivity was 0.1~Jy.
They found that the median angular size of the resolved cores is about 0.26 mas
and a significant fraction of the sources have a source frame
core brightness temperature in excess of $10^{12}$~K
based on the imaging and model-fitting with Gaussian models
for the compact radio jets.
A lower limit on the brightness temperature of the VLBI cores was determined
using the Difwrap software package.

\subsection{Analysis\label{analysis}}

\subsubsection{Intrinsic Brightness Temperatures\label{analysis-To}}

The observed brightness temperatures for a sample of compact radio jets,
compiled from VLBI surveys at different frequencies,
can be used to estimate their intrinsic brightness
temperature at each corresponding frequency.
The intrinsic brightness temperature $T_{0}$ is 
one of the intrinsic physical properties of the relativistic jets
together with the Lorentz factor $\gamma_{\rm j}$ and
the angle to the line of sight $\theta_{\rm j}$.
These intrinsic parameters can be related with the observed properties
of the jets such as 
the apparent jet speed $\beta_{\rm app}$,
the flux density and size of the emitting regions,
and hence the observed brightness temperature $T_{\rm b}$ as
in \cite{lee13}:
   \begin{equation}
   \label{eqn:Tb-1}
   \delta = \frac{1}{\gamma_{\rm j} (1-\beta {\rm cos}\theta_{\rm j})},
   \end{equation}
   \begin{equation}
   \label{eqn:Tb-2}
   \beta_{\rm app} = \frac{\beta {\rm sin}\theta_{\rm j}}{1-\beta {\rm cos}\theta_{\rm j}},
   \end{equation}
and
   \begin{equation}
%  T_{\rm b} = T_{\rm 0} \delta^{n},
   T_{\rm b} = T_{\rm 0} \delta,
   \label{eqn:Tb-3}
   \end{equation}
where $\delta$ is the Doppler factor and
$\beta = (1-{\gamma_{\rm j}}^{-2})^{1/2}$ 
is the speed of the jet in the rest frame
of the source (in units of $c$).
If we assume that (a) the compact radio jets are narrow and straight
with no bends as an ideal relativistic jet,
(b) the maximum apparent jet speed is the same as the jet flow speed,
(c) all jets in the sample have the same intrinsic brightness temperature
$T_{0}$ at a given observing frequency
and $T_{0}$ does not evolve with redshift,
and (d) the jets are at the critical viewing angle
$\theta_{\rm c}={\rm arccos}~\beta$ for the maximal apparent
jet speed at a given $\beta$,
then the observed brightness temperature can be related to the maximum
jet speed as again in \cite{lee13}:
   \begin{equation}
   \label{eqn:delta}
   \delta \simeq \beta_{\rm app}
   \end{equation}
and
   \begin{equation}
   \label{eqn:Tb-Bapp-To}
   T_{\rm b} \simeq \beta_{\rm app} T_{\rm 0}. 
   \end{equation}
This simple relation between the observed
brightness temperature and the apparent maximum jet
speed is well described as the solid line in Figure 1 of \cite{lee13}.
Using equations (\ref{eqn:Tb-1}) and (\ref{eqn:Tb-2}),
the apparent jet speed $\beta_{\rm app}$ can be related
to the Doppler factor $\delta$
and the Lorentz factor $\gamma_{\rm j}$:
   \begin{equation}
   \label{eqn:Tb-4}
   \beta_{\rm app} = \sqrt{ (\delta \gamma_{\rm j} \beta)^2 - (\delta \gamma_{\rm j} - 1)^2 }.
   \end{equation}
for the maximum and minimum possible Doppler factors 
$\delta_{\rm max} = 1/\gamma_{\rm j}$ and 
$\delta_{\rm min} = 1/(\gamma_{\rm j} - \sqrt{{\gamma_{\rm j}}^2 -1} )$.
This relation shows
the apparent speeds as a function of $T_{\rm b}$
for jets with Lorentz factor
of $\gamma_{\rm j}$,
and appears as an evolope in the $\beta_{\rm app}-T_{\rm b}$
plane~\citep[and as three solid lines in Figure 1 of][]{lee13}.
According to \cite{hom+06},
the intrinsic brightness temperature is derived from
the curve $\beta_{\rm app}(T_{\rm b})=T_{\rm b}/T_{0}$
which passes through the maximum of the envelope (equation \ref{eqn:Tb-4})
defined by the maximum apparent speeds as a function of
observed brightness temperature.
This curve approximately divides the data in
the $\beta_{\rm app}-T_{\rm b}$ plane
such that about 25\% of the data are located above the curve.
Under the assumptions (a)-(d) for the sample of compact radio jets,
the derived value of $T_{0}$ can represent the intrinsic brightness temperature
of the compact radio jets with an uncertainty range.
The uncertainty range of $T_{0}$ may be determined between $T_{65\%}$
and $T_{85\%}$, where $T_{65\%}$ is $T_{0}$ for 35\% of the jets above
the curve, and $T_{85\%}$ is for 15\% of the jets above the curve.

\begin{figure*}[!t]
        \centering 
	\includegraphics[trim=4mm 0mm 1mm 1mm, clip, width=173mm]{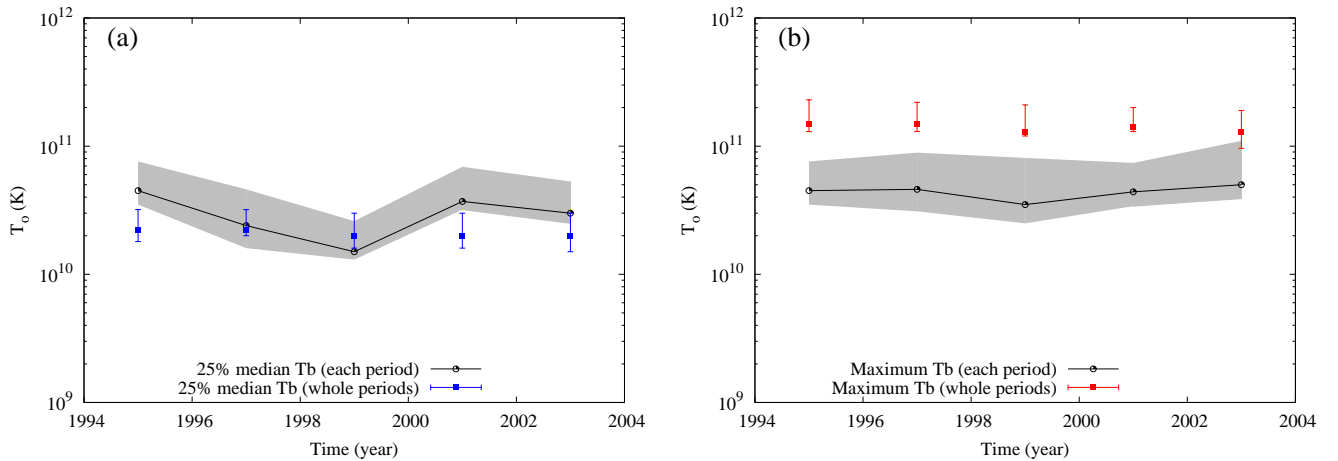}
        \caption{ 
	Variability of the derived intrinsic brightness temperatures $T_{0}$
	at 15~GHz
	(black dot) based on (a) the 25\% median $T_{\rm b}$
	and (b) the maximum $T_{\rm b}$ for each epoch.
	The $T_{0}$ for the corresponding sample of two-year time period
	derived using
	the observations over the whole period of 1994-2003
	are shown (a) in blue dots for the 25\% median and
	(b) in red dots for the maximum.
	The uncertainty range (grey area and errorbar) of $T_{0}$ is
	defined with $T_{0}$ for the 65\% and 85\% fractions. 
        }
        \label{fig1} 
\end{figure*}

\subsubsection{Brightness Temperature vs. Time\label{analysis-time}}

Variability of the compact radio jets in flux density plays
a critical role in determining the representative intrinsic properties,
when one investigates the multifrequency intrinsic brightness temperatures
based on independent VLBI surveys.
Multiepoch VLBI observations over a long period (e.g., 10 years)
for a sample of variable compact jets
may provide us with a chance 
to observe the compact jets in their typical low brightness state.
The typical low brightness state can be defined 
by selecting the median of the lower half of the brightness
temperatures for a compact jet called as ``25\% median''~\citep{hom+06}.
However, some VLBI surveys have been conducted over a relatively
short period of time (e.g., a few years for the 86~GHz VLBI survey). 
Although a compact jet has been observed in several times
during such a short period, the 25\% median of brightness temperatures
may be different from those with longer-period observations.
Therefore it is important to investigate  
the influence of the source variability on determining
the intrinsic brightness temperature.

Using the 15~GHz VLBA observations conducted in
1994-2003~\citep[used by][for their analysis]{hom+06},
we investigated the intrinsic brightness temperature
between different epochs 
of observations spanning two years, between 1994 and 2003.  
To select the compact jets observed in each epoch
we required that they also be observed in the 86~GHz VLBI survey.
The number of sources selected are 
68 for 1994-1995, 76 for 1996-1997, 75 for 1998-1999,
66 for 2000-2001, and 84 for 2002-2003.
The sample sizes are different for each two-year period,
and relatively smaller than those
used by \cite{hom+06} and \cite{lee13}. This may cause 
uncertainty in the investigation.
However, the smaller sample size may not play a critical role in
reliable determination of the intrinsic brightness temperature
for each epoch since we found similar intrinsic brightness temperatures
for different samples of the compact jets
based on their brightness temperatures over the whole period
of 1994-2003 (see below). 
The number of observations for one source over a period of two years
is in a range of 1-5. For the time period of 1994-1996, all sources were
observed once, and for other time periods, some sources were observed
several times. 
The mean number of observations for one source over a single two-year period
is 1.6. 
We determined the 25\% median and maximum of the brightness temperatures
as a function of time for individual sources
observed more than once per time period.
For those sources observed only one time,
there is no 25\% median or maximum.
Therefore, we used the single measurement of brightness temperature
in the following analysis.
For those time-dependent sources we estimated the corresponding intrinsic
brightness temperatures and found that
they are variable for the 25\% median case (Figure~\ref{fig1}a),
whereas less variable for the maximum case (Figure~\ref{fig1}b).
The measured intrinsic brightness temperatures are in ranges
of $T_{0}=1.5\times10^{10} \sim 4.5\times10^{10}$~K for the 25\% median case,
and of $T_{0}=3.5\times10^{10} \sim 5.0\times10^{10}$~K for the maximum case.
Moreover,
from 1998-1999 to 2000-2001, the intrinsic brightness temperature
for the 25\% median case has increased by a factor of ~2.5.
The difference in the estimation of the intrinsic brightness temperature
may be attributed to the different characteristics of each sample.
In order to see if the difference of the sample in size affects
the determination of the intrinsic brightness temperature,
we determined the 25\% median and maximum of the jet brightness temperature
as a function of time for individual sources
in the sub-sample,
based on their brightness temperatures obtained over the whole period
of 1994-2003.
We found that the characteristic intrinsic brightness
temperatures for the sub-samples are within very small ranges of
$T_{0}=2.0\times10^{10} \sim 2.2\times10^{10}$~K for the 25\% median case,
and $T_{0}=1.3\times10^{11} \sim 1.5\times10^{11}$~K for the maximum case,
implying the difference of
the sub-samples in size may not affect strongly the determination 
of the intrinsic brightness temperature of each sub-sample.

By investigating the observed brightness temperatures at 15~GHz 
in multiple epochs, we found that the determination of
the intrinsic brightness temperature is affected by 
the variability of individual jets in flux density at 
the time scales of a few years.
Therefore, for the multifrequency statistical study of the intrinsic
properties of the compact jets, it is important to 
use contemporaneous VLBI observations by conducting
multifrequency observations within narrow period of time or
by conducting simultaneous observations at multiple frequencies.

\subsubsection{Brightness Temperature vs. Frequency\label{analysis-freq}}

Assuming the maximum jet speed observed at 15~GHz \citep{lis+09,lis+13}
represents the flow speed of the jets, \cite{lee13} applied 
the statistical method described in Section~\ref{analysis-To}
to the sample of brightness temperatures
derived at 86~GHz for which 15~GHz brightness temperatures
were available. They found that the derived intrinsic brightness
temperature for 98 compact radio jets at 86~GHz was 
significantly lower than the one at 15~GHz.
They also found that the estimated Doppler factors
from the derived intrinsic temperatures are significantly lower
for the 86~GHz VLBI cores than for the 15~GHz cores.
This has shown that multifrequency VLBI survey observations
together with the above statistical method
can be used to determine the intrinsic brightness temperatures
representing the intrinsic properties of different regions
of the compact radio jets.

In order to investigate the intrinsic brightness temperature
of compact radio jets as a function of frequency,
we compiled the observed brightness temperatures of the compact jets
from VLBI surveys conducted at 2, 5, 8, 15, and 86~GHz. 
For maintaining a contemporaneity of the multifrequency samples
we selected the observed brightness temperatures measured
in 2001-2003, which includes the period of the 86~GHz VLBI observations.
We excluded the compact jets which were not observed in the 86~GHz VLBI
survey~\citep{lee+08}.
The number of compact jets selected for this study is
59 for the 2/8~GHz, 93 for the 15~GHz, and 98 for the 86~GHz observations.
Since we found no suitable jets observed at 5~GHz in 2001-2003,
we used the 5~GHz VSOP observations in 1996 and selected 65 jets.
We investigated the redshift distribution of the selected sources,
and found no strong correlation of the observed brightness temperatures
with their redshift. 
By using the same criteria as those
adopted in \cite{hom+06} and \cite{lee13},
we found that
the characteristic intrinsic brightness temperatures of each sample are 
$T_{0}=2.0\times10^{10}$~K at 2~GHz,
$T_{0}=3.3\times10^{10}$~K at 5~GHz,
$T_{0}=4.3\times10^{10}$~K at 8~GHz,
$T_{0}=3.4\times10^{10}$~K at 15~GHz,
and $T_{0}=4.8\times10^{9}$~K at 86~GHz.

\section{Results\label{sec:results}}

\begin{figure}[!t]
        \centering 
	\includegraphics[trim=4mm 0mm 1mm 1mm, clip, width=83mm]{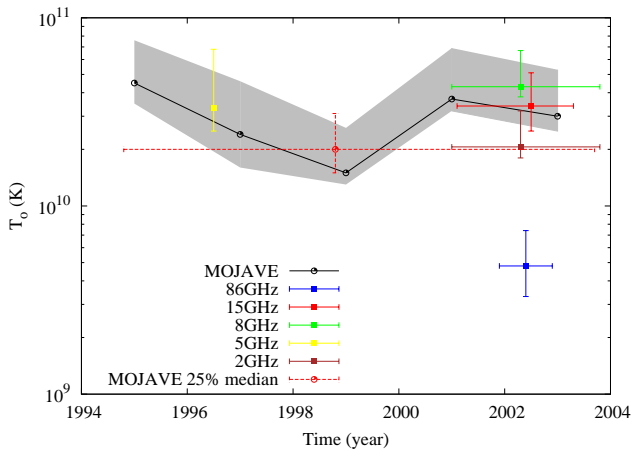}
        \caption{ 
	Determined $T_{0}$ for our sample at 2~GHz (brown square),
	5~GHz (yellow square), 8~GHz (green square), 15~GHz (red square),
	and 86~GHz (blue square) are plotted at their median date with
	the range of the epoch. The 25\%-median $T_{0}$ for the 15~GHz
	observations in 1994-2003 (MOJAVE) is shown with a red circle. 
	Variability of the intrinsic brightness temperatures at 15~GHz
	(black dot) are also plotted for comparison.
	The uncertainty range (grey area and errorbar) of $T_{0}$ is
	determined at the 65\% and 85\% fractions. 
        }
        \label{fig2} 
\end{figure}

\begin{figure}[!t]
        \centering 
	\includegraphics[trim=4mm 0mm 1mm 1mm, clip, width=83mm]{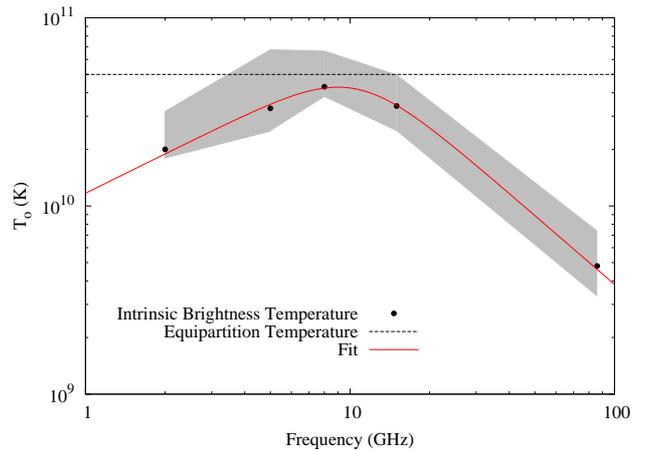}
        \caption{ 
	Intrinsic brightness temperatures of our sample at 2-86~GHz
	(black dot). The uncertainty range (grey area) of $T_{0}$ is
	defined at the 65\% and 85\% fractions. 
	The red solid line shows the best fit to the data with the smooth broken
	power-law as described in the text.
	The dashed line indicates the equipartition temperature.
        }
        \label{fig3} 
\end{figure}

The determined intrinsic brightness temperatures for
the multifrequency samples are different from each other
as shown in Figure~\ref{fig2}, 
implying that
the compact radio jets in our sample yield
different characteristic brightness temperatures for  
the corresponding emission regions (VLBI core)
at different frequencies.
The intrinsic brightness temperature at 15~GHz for our sample (red dot)
is different from the one calculated over
the whole period of the 15~GHz observations
(red circle) due to the variability effect.
The intrinsic brightness temperatures at 86~GHz is much lower than  
those at other frequencies. 
The 8~GHz sample gives the highest brightest temperature of
$T_{0}=4.3\times10^{10}$~K which is very close to
the equipartition temperature of $T_{0}=5\times10^{10}$~K,
indicating that the jet emission regions at 2--86~GHz are in
magnetically dominated environments.
We have also fitted the multifrequency intrinsic brightness temperatures
with a smoothly broken power-law function
   \begin{equation}
   \label{eqn:powerlaw}
   T_{0}(\nu) = T_{\rm 0,j}\left[\left(\frac{\nu}{\nu_{\rm j}}\right)^{\alpha_1 n}
           + \left(\frac{\nu}{\nu_{\rm j}}\right)^{\alpha_2 n}\right] ^{-1/n},
   \end{equation}
where $T_{\rm 0,j}$ is the intrinsic brightness temperature at
an observing frequency $\nu_{\rm j}$,
$-\alpha_1$ and $-\alpha_2$ are the slopes at higher and lower frequencies
than the observing frequency, respectively,
and $n$ is a numerical factor governing the sharpness of the break.
The broken power-law function peaks at a critical frequency $\nu_{\rm c}$
which depends on the numerical factor $n$ and the best fitting parameters,
$\nu_{\rm j}$, $\alpha_1$ and $\alpha_2$ as
$\nu_{\rm c}=(-\alpha_2/\alpha_1)^{1/n(\alpha_1-\alpha_2)}$,
providing the peak brightness temperature
value $T_{\rm 0,c}=T_{0}(\nu_{\rm c})$.
As shown in Figure~\ref{fig3}, 
the best fitting was obtained for 
$T_{\rm 0,j}=(5.97\pm0.24)\times10^{10}$~K,
$\nu_{\rm j}=10.29\pm0.68$~GHz,
$\alpha_1=1.19\pm0.03$,
and
$\alpha_2=-0.67\pm0.06$,
after choosing the numerical factor $n=1.97$,
using an implementation of the nonlinear least-squares
Marquardt-Levenberg algorithm.
From the best fitting parameters, we found that
the intrinsic brightness temperature $T_{\rm 0}$ increases
as $T_{\rm 0}\propto\nu^{\epsilon}$ with $\epsilon\approx0.7$ below
the critical frequency $\nu_{\rm c}\approx9~{\rm GHz}$
where energy losses begin to dominate the emission,
and above $\nu_{\rm c}$, $T_{\rm 0}$ decreases
with $\epsilon\approx-1.2$.
We also found that the peak value of the fitted curve,
$T_{\rm 0,c}\approx3.4\times10^{10}$~K,
is close to the equipartition temperature, implying that
the VLBI cores visible at 2-86~GHz may be representing
jet regions where the magnetic field energy dominates
the total energy in jets.

\section{Discussion\label{sec:dis}}

\begin{figure}[!t]
        \centering 
	\includegraphics[trim=6mm 4mm 2mm 3mm, clip, width=83mm]{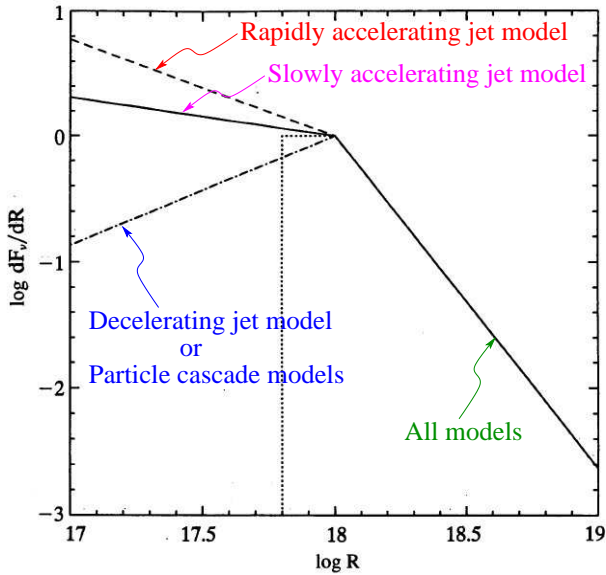}
        \caption{ 
	Prediction for flux density per 
	unit length of relativistic jets
	for various jet models:
	decelerating jet model or particle cascade model (dot dashed line),
	slowly accelerating jet model (solid line below $R=10^{18}$~cm, and
	rapidly accelerating jet model (dashed line).
	Above $R=10^{18}$~cm, all models show the same evolution of the flux density. The flux density is in arbitrary units~\citep[taken from][]{mar95}.
        }
        \label{fig4} 
\end{figure}

\cite{mar95} predicted that the flux density per unit length
of a relativistic jet $dF_{\nu}/dR$ evolves along the distance from the origin
(or from the central engine of jets) $R$
as $dF_{\nu}/dR \propto R^{-\xi}$
showing a power-law slope of $\xi=2.6$ in the outer region
($R > R_{\rm c}$) of jets
and varying slopes of $-1$ to $+1$ in the inner region ($R < R_{\rm c}$)
depending on the jet model assumed.
For the decelerating jet model or the particle cascade model,
$\xi\sim -1$,
and for the rapidly accelerating jet model, 
$\xi\sim +1$
as distance $R$ increases from the central engine (Figure~\ref{fig4}).
One may relate the prediction of flux density 
$dF_{\nu}/dR$ as a function of distance
with the intrinsic brightness temperature
$T_{0}$ 
as a function of observing frequency for the VLBI core region
based on the following argumentation. 
The VLBI core, the most compact feature in VLBI images,
is believed to be located in the jet regions 
where the optical depth is unity.
Therefore, the distance of the core
$R_{\rm core}$
from the central engine
should depend on the obverving frequency $\nu_{\rm obs}$,
in the form
$R_{\rm core}\propto\nu_{\rm obs}^{-1/k_{\rm r}}$~\citep{kon81}.
The power-law index $k_{\rm r}$ varies depending on
the distribution of the electron energy
and on the magnetic field and particle density distributions.
For the flat-spectrum cores under equipartition condition,
a general prediction is $k_{\rm r}\sim1$~\citep{kon81,lob98}. 
Since VLBI observations generally show that the size of the compact jet
has a power-law dependence on distance as $\theta_{\rm j}\propto R^l$,
the jet size should be related to the observing frequency
as $\theta_{\rm j}\propto \nu^{-l/k_{\rm r}}$.
The power-law index $l$ also depends on the physical properties of jets.
\cite{PK12} found $0.8<l<1.2$ from the the jet observations at 2 and 8~GHz.
If we take the average value of $l=1$ and the general prediction of
$k_{\rm r}\sim1$,
the factor $\nu_{\rm obs}^2 \Omega$ is constant
along the jet axis, where $\Omega$ is the solid angle covering
the section of the jets,
and hence $dF_{\nu}/dR \propto T_{0}/(\nu_{\rm obs}^2 \Omega)\propto T_{0}$.
Therefore, it is implied that
$T_{0} \propto R^{-\xi} \propto \nu_{\rm obs}^{\xi}$ with $\xi = +2.6$
below a critical frequency $\nu_{\rm c}$, which corresponds to 
the peak frequency of the spectrum of the jet out to distance $R_{\rm c}$, and
with $-1 < \xi < +1$ beyond $\nu_{\rm c}$, depending on the jet models:
$\xi\sim-1$ for the decelerating jet model and $\xi\sim+1$ for the rapidly
accelerating jet model.
Our result, $T_{0}\propto\nu_{\rm obs}^{-1.2}$ ($\nu_{\rm obs}>9$~GHz)
and hence $\xi=-1.2$,
supports the decelerating jet model or particle cascade model  
for the sample of the compact jets.
At $\nu_{\rm obs}<9$~GHz,
our result, $T_{0}\propto\nu_{\rm obs}^{0.7}$
and hence $\xi=+0.7$,
may not be fully consistent with the prediction.
Taking into account the uncertainty of the best fit,
$\Delta\xi=0.06$, the difference between the theory and observation
($\xi=0.7\pm0.06~{\rm vs.}~2.6$) is indeed significant.
This is potentially very interesting and could provide a serious test
of the \cite{mar95} model - which would be a truly new result.

\section{Conclusions\label{sec:con}}

We investigated the brightness temperatures of
98 compact jets obtained from high resolution VLBI observations
at 2, 5, 8, 15, and 86~GHz. 
Based on the analysis of the 15~GHz observations at multiple epochs,
we found that the determination of
the intrinsic brightness temperature for our sample is affected by 
variability of individual jets in flux density over
the time scales of a few years.
Therefore, it is important to use contemporaneous
VLBI observations for the multi-frequency analysis
of the intrinsic brightness temperatures.
Since we were able to compile
the high resolution VLBI observations conducted in 2001-2003,
the results should not be strongly affected by the flux density variability.
The analysis with the contemporaneous multifrequency VLBI observations shows
that the intrinsic brightness temperature $T_{\rm 0}$ increases
as $T_{\rm 0}\propto\nu_{\rm obs}^{\xi}$ with $\xi=+0.7$ below
the critical frequency $\nu_{\rm c}\approx9~{\rm GHz}$
and above $\nu_{\rm c}$, $T_{\rm 0}$ decreases
with $\xi=-1.2$.
This result supports the decelerating jet model.
We also found that the characteristic intrinsic brightness temperatures
at 2--86~GHz are lower than the equipartition temperature and 
the peak value of $T_{\rm 0,c}\approx3.4\times10^{10}$~K
is close to the equipartition temperature, implying that
the VLBI cores observable at 2--86~GHz may be representing
jet regions where the magnetic field energy dominates
the total energy in jets.

%%% ACKNOWLEDGMENTS (IF ANY) %%%%%%%%%%%%%%%%%%%%%%%%%%%%%%%%%%%%%%%%

\acknowledgments

I would like to thank the anonymous referee for important comments
and suggestions, which have enormously improved the manuscript.
I am grateful to Andrei Lobanov for carefully reading and kindly commenting
on the manuscript.
The VLBA is an instrument of the National Radio Astronomy Observatory,
which is a facility of the National Science Foundation operated under
cooperative agreement by Associated Universities, Inc.

\end{document}